\documentstyle[aps,prl,preprint,floats,epsfig]{revtex}

\def\LA{\Lambda}
\def\To{\rightarrow}
\def\CASMIN{\Xi^- \To \LA \pi^-}

\def\SLOPEXA{-0.291 \pm 0.021\mbox{(stat)}^{+0.019}_{-0.015}\mbox{(syst)}}
\def\ACP{-0.057 \pm 0.064 \pm 0.039 }

\begin{document}

\draft  

\preprint{\tighten\vbox{\hbox{\hfil CLNS 98/1587}
                        \hbox{\hfil CLEO 98-16}
}}

\title{Search for Direct CP Violation in $\Xi$ Hyperon Decay}  

\author{CLEO Collaboration}
\date{\today}

\maketitle
\tighten

\begin{abstract} 
  Using data collected with the CLEO II detector we have performed a
  search for direct CP violation in the $\Xi$ hyperon system.  CP
  violation gives rise to an asymmetry, ${\mathcal{A}}$, between the
  parity-violating angular distributions of the decay chains $\Xi^- \To
  \LA \pi^-, \ \LA \To p \pi^-$ and $\bar{\Xi}^+ \To \bar{\LA} \pi^+,
  \ \bar{\LA} \To \bar{p} \pi^+$.  In the Standard Model, ${\mathcal{A}}
  \approx 10^{-4}$ to $10^{-5}$.  If CP violation were found at a higher
  level it could indicate the presence of new physics.  We find no
  evidence for CP violation: ${\mathcal{A}} = \ACP$.  We also obtain
  $A_{\Xi} = -0.070 \pm 0.064 \pm 0.045$ and
  $\alpha_{\bar{\Xi}}=0.49^{+0.06}_{-0.05}\pm0.10$.
\end{abstract}
\pacs{PACS numbers: 14.20.Lq,14.20.Jn,14.65.Dw,13.30.-a,13.30.Eg}
\newpage

{
\renewcommand{\thefootnote}{\fnsymbol{footnote}}

\begin{center}
D.~E.~Jaffe,$^{1}$ G.~Masek,$^{1}$ H.~P.~Paar,$^{1}$
E.~M.~Potter,$^{1}$ S.~Prell,$^{1}$ V.~Sharma,$^{1}$
D.~M.~Asner,$^{2}$ A.~Eppich,$^{2}$ J.~Gronberg,$^{2}$
T.~S.~Hill,$^{2}$
% ...
\end{center}
 
\small
\begin{center}
$^{1}${University of California, San Diego, La Jolla, California 92093}\\
$^{2}${University of California, Santa Barbara, California 93106}\\
% ...
\end{center}
 
\setcounter{footnote}{0}
}
\newpage

To date $CP$ violation has only been observed in the neutral kaon
system~\cite{CRONIN} and its origin remains unknown.  It is believed
that there is insufficient $CP$ violation in the minimal standard model
(MSM) to generate the matter-antimatter asymmetry of the
universe~\cite{KOLB}. Searches for additional $CP$ violation beyond the
MSM may help reconcile this problem.

In the MSM $CP$ violation effects are due to a single complex phase in
the CKM quark mixing matrix~\cite{CKM}. In the standard phase
convention the two matrix elements with large phases are $V_{ub}$
and $V_{td}$. Because these elements have small magnitudes and involve 
the third generation, $CP$ violation in kaon decays is small. In the kaon 
system, $\epsilon$ is a measure of CP violation due to mixing~\cite{BRAN}
and $\epsilon'$ is a measure of CP violation in the decay amplitude, 
called direct CP violation. Theoretical errors are too large to interptret 
the measured value of $\epsilon'/\epsilon$ as evidence for the MSM mechanism
of CP violation or for physics beyond the MSM. In the MSM, and its extensions, 
large $CP$-violating effects are anticipated in $B$ meson decays. $CP$ 
violation may also be observed by comparing 
$\Delta S = 1$ hyperon and antihyperon non-leptonic decays~\cite{CHAU,PAK}.  
The latter is the subject of this Letter.
 

In hyperon and kaon decay gluonic penguin transitions~\cite{SHIF} are
thought to give rise to the relative weak phase difference between
particle and antiparticle amplitudes. In kaon decays the interference
is between the two isospin amplitudes, whereas in $\Xi$ decay it is
between the $S$ and $P$ wave amplitudes. Consequently, in $\Xi$ decay
the $CP$ violation observable is not a rate difference but a difference,
${\cal A}$, in the degree of parity violation in charge conjugate
states. MSM predictions of
${\cal A}$ depend on the values of $\epsilon$, the top quark mass and
the hadronic matrix element. The dominant uncertainty is due to
incomplete knowledge of the hadronic matrix element.  The Superweak
model~\cite{WOLF} and models with a very heavy neutral Higgs where there are no
$|\Delta S|=1$ $CP$-odd effects predict no $CP$ asymmetries. Other
models in which $|\Delta S|=1$ $CP$
nonconservation is dominant, predict asymmetries which are on the order
of, or larger than those in the MSM.  Theoretical predictions of
$A_{\Xi}$ and $A_{\LA}$ range from $10^{-4}$ to $10^{-5}$
~\cite{XGHE2,DONOG,XGHE1}.

Parity violation occurs in hyperon ${1 \over 2}^+ \To {1 \over 2}^+ 0$
decays due to the existence of two orbital angular momentum amplitudes
of opposite parity. The parity violation observable is an asymmetry in
the angular decay distribution due to interference between the two
amplitudes.  In the decay $\CASMIN$ followed by $\LA \To p \pi^-$, the
$\LA$ is produced with a polarization equal to
\begin{equation}
 {\bf P_{\LA}} = { (\alpha_\Xi + \hat{\bf \LA} \cdot {\bf P_\Xi })\hat{\bf \LA}
- \beta_\Xi ( \hat{\bf \LA} \times {\bf P_\Xi} ) - \gamma_\Xi \hat{\bf \LA}
\times ( \hat{\bf \LA} \times {\bf P_\Xi} ) \over 
( 1 + \alpha_\Xi \hat{\bf \LA}
\cdot
{\bf P_\Xi} ) }
\label{polarize}
\end{equation}
where {\boldmath $P_{\Xi}$} is the $\Xi$ polarization, $\alpha_{\Xi}
,~\beta_{\Xi}$ and $~\gamma_{\Xi}$ are the $\Xi$ asymmetry parameters,
which measure the degree of parity violation in $\Xi$ decay, and ${\bf
  \hat{\LA}}$ is a unit vector along the $\LA$ momentum in the $\Xi$
rest frame~\cite{KALLEN}. If the $\Xi$ polarization is unobserved, or if
the $\Xi$ is not polarized, Equation ~\ref{polarize} reduces to ${\bf
  P_{\LA}} = \alpha_{\Xi} {\bf \hat\LA }$.  The angular distribution of
the proton from the decay of the $\LA$ is therefore
\begin{equation}
{dN \over d \cos \theta_{\LA}} \propto 1 + \alpha_{\Xi} \alpha_{\LA}
\cos \theta_{\LA}
\label{xiasym}
\end{equation}
where $\theta_{\LA}$ is the angle
between the proton momentum vector in the $\LA$ rest frame
and ${\bf \hat \LA}$,
and $\alpha_{\LA}$ is the $\LA$ decay asymmetry which 
measures the degree of parity violation in $\LA$ decay.
If $CP$ is conserved
in the decays $\LA \To p \pi$ and $\Xi
\To \LA \pi$, $\alpha_{\LA} = -\alpha_{\bar{\LA}}$
and $\alpha_{\Xi} = -\alpha_{\bar{\Xi}}$, respectively.  Therefore the
$CP$-violating asymmetry parameters, $A_{\LA}$ and $A_{\Xi}$, are
defined as ~\cite{PAK}
\begin{equation}
A_{\LA} = { \alpha_{\LA} + \alpha_{\bar{\LA}}
\over \alpha_{\LA} - \alpha_{\bar{\LA}}} \ \ {\rm and} \ \
A_{\Xi} = { \alpha_{\Xi} + \alpha_{\bar{\Xi}}
\over \alpha_{\Xi} - \alpha_{\bar{\Xi}}}.
\end{equation}
If CP is conserved in the decay sequence $\Xi \To \LA \pi, \ \LA \To p \pi$,
$\alpha_{\Xi} \alpha_{\LA}= \alpha_{\bar{\Xi}}
\alpha_{\bar{\LA}}$. Therefore we measure the $CP$-violating asymmetry
parameter~\cite{E871}:
\begin{equation}
{\mathcal{A}} = { \alpha_{\Xi} \alpha_{\LA} - \alpha_{\bar{\Xi}}
\alpha_{\bar{\LA}}
\over \alpha_{\Xi} \alpha_{\LA} + \alpha_{\bar{\Xi}} \alpha_{\bar{\LA}}}
= A_{\Xi} + A_{\LA}  + {\mathcal{O}}(\Delta \alpha_{\Xi} \Delta
\alpha_{\LA})
\simeq A_{\Xi} + A_{\LA}
\label{cpasym}
\end{equation}
where $\Delta \alpha_{\Xi, \LA} = \alpha_{\Xi, \LA} + \alpha_{\bar{\Xi},
\bar{\LA}}$.

The data sample in this study was collected with the CLEO II
detector~\cite{KUBOTA} at the Cornell Electron Storage Ring (CESR). The
integrated luminosity consists of 4.83 $fb^{-1}$ taken at and just below
the $\Upsilon (4S)$ resonance, corresponding to approximately 5 million
$e^+ e^- \rightarrow c\overline{c}$ and 1.25 million $e^+e^-\rightarrow
s\overline{s}$ events. The latter process is the dominant source of
$\Xi$'s in this analysis.


The $\LA$ is reconstructed in the $p \pi^{-}$ decay mode\footnote{Charge conjugation 
is implied in this paper except were explicitly noted}. We require two
oppositely charged tracks to originate from a common vertex.
The positive track is
required to be consistent with a proton hypothesis\footnote{Hadronic particles are 
identified by requiring specific ionization energy loss measurements ($dE/dx$), combined 
with time-of-flight (TOF) information when available. The two measurements are combined 
into a joint probability for the particle 
to be a pion, a kaon or a proton. A charged track is defined to be consistent with a
particle hypothesis if its probability is greater than 0.003.}.
The momentum of the
$\LA$ candidate is calculated by extrapolating the charged track momenta
to the secondary vertex.  The invariant mass of $\LA$ candidates is
required to be within three standard deviations (3$\sigma$ = 6.0
MeV/c$^2$) of the known $\LA$ mass.  Track combinations which
satisfy interpretation as $K_s^0 \To \pi^+ \pi^-$ are rejected.
Combinatoric background is reduced by requiring the momentum of the $\LA$ candidates
to be greater than 800 MeV/c. Due to this requirement $B$ decays are not
a signifcant source of $\Xi$'s.

\begin{figure}[!h]
\centerline{\psfig{figure=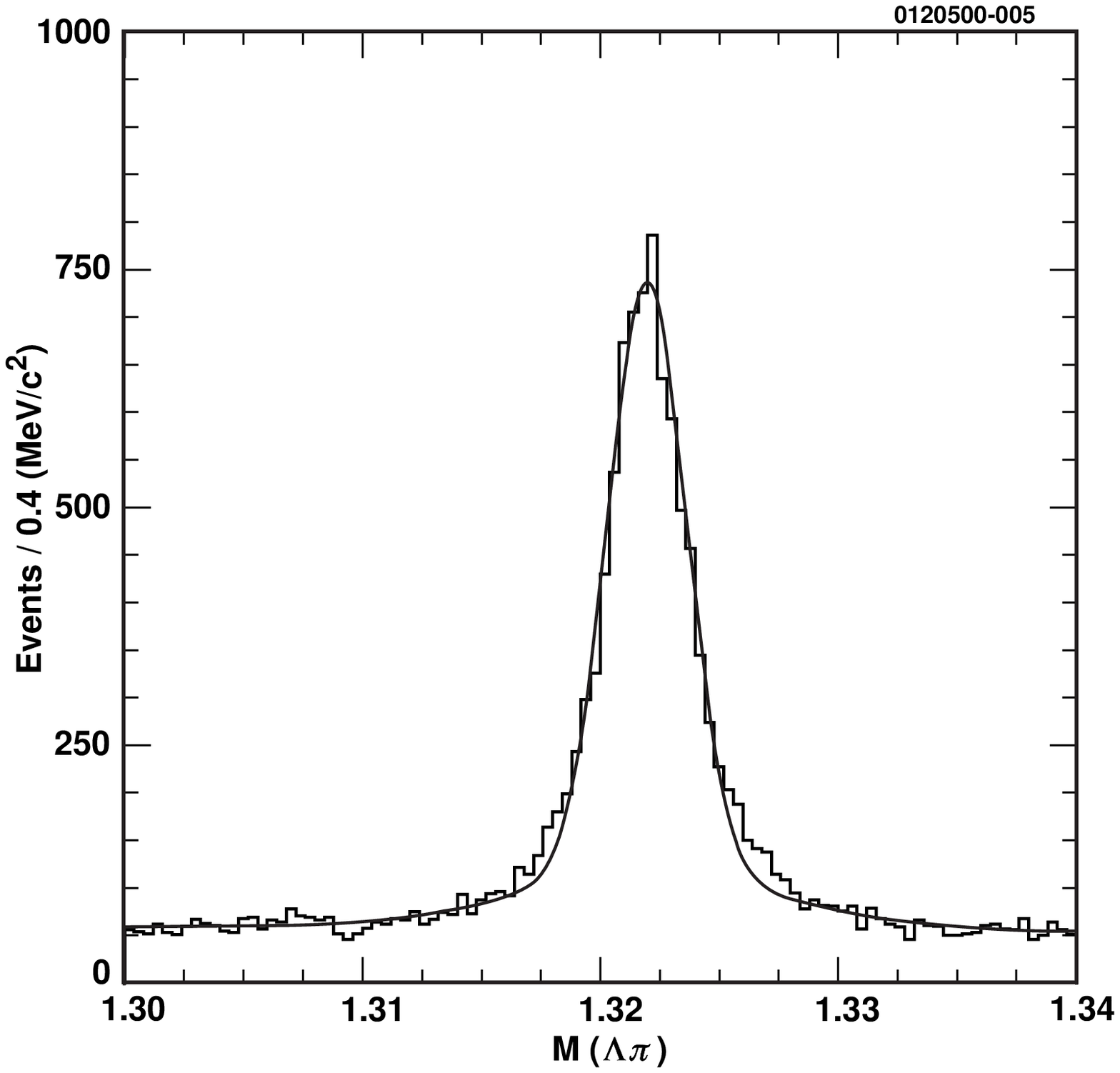,height=0.4\textwidth}}
\caption{ $\LA \pi^-$ invariant mass distribution.}
\label{Xi9}
\end{figure}

The $\Xi^-$ is reconstructed in the $\LA \pi^-$ decay mode.  $\Xi^{-}$
candidates are formed by combining each $\LA$ candidate with a
negatively charged track consistent with a pion hypothesis.  The $\Xi^-$
candidate vertex is formed from the intersection of the $\LA$ momentum
vector and the negatively charged track.  To obtain the $\Xi^-$
momentum, and $\LA \pi$ invariant mass, the momentum of the charged track is recalculated at the new
vertex.  The invariant $\LA \pi$ mass is shown in  Fig. ~\ref{Xi9} 
for all $\Xi^-$ candidates satisfying the selection criteria.

The $\LA \pi$ mass is fit is to a double Gaussian with 
widths and ratio of areas fixed from a
GEANT~\cite{GEANT} based Monte Carlo (MC) simulation of the detector and
a first order Chebyshev polynomial to describe the combinatorial
background.  The $\Xi \rightarrow \LA \pi$ events are generated using
LUND/JETSET 7.3~\cite{LUND}, in the process $e^+ e^- \rightarrow s\bar{
  s} \rightarrow \Xi X$.  The MC momentum distributions and production
angles of all particles in the decay chain are in good agreement with
observation.  We find $8434 \pm 109$ events consistent with $\Xi
\rightarrow \LA \pi$.  The mean of the $\LA \pi$ invariant mass
distribution is in agreement with the MC simulation.

The $\cos\theta_{\LA}$ resolution function 
has both Gaussian and symmetric non-Gaussian components.
The resolution in $\cos \theta_{\LA}$, $\sigma_{av}$, defined to be the
average of the rms variance of the two components, weighted by their 
relative normalizations
is $\sigma_{av} = 0.030$.

The product of the decay asymmetry parameters,
$\alpha_{\Xi}\alpha_{\LA}$, is measured using a one dimensional unbinned
maximum likelihood fit to the $\cos \theta_{\LA}$ distribution given in
Equation ~\ref{xiasym}, in a manner similar to ~\cite{SCHMIDT}.  This
technique enables a likelihood fit to be performed to
variables modified by experimental acceptance and resolution. 
The probability function of the signal $\Gamma_s$ is determined 
by generating one high statistics MC sample of $\CASMIN$ at 
fixed $\alpha_{\Xi}\alpha_{\LA}$.   By suitable weighting of the
accepted MC events a
likelihood is evaluated for each data event for trial values of
$\alpha_{\LA}\alpha_{\Xi}$, and a fit performed.  The probability for
each event is determined by sampling $\Gamma_S$ using an interval
centered on each data point.


Background is incorporated directly into the fit by constructing the
log-likelihood function:
$
\ln {\mathcal{L}} = \sum_{i=1}^{N} ln (
P_S \Gamma_S +
P_B \Gamma_B )
$
where $N$ is the number of events in the signal region,
defined to be within 7.5 Mev/c$^2$ ($ \pm 3 \sigma$) of the $\Xi$
mass, and $P_S$ and
$P_B$ are the probabilities that events are signal and background
respectively. 
The probability distribution of background in the signal region,
$\Gamma_B$, is determined from 
15 MeV/c$^2$ wide ($\LA \pi$) mass sidebands above and below the signal
region.

To ensure that there is no gross bias in the analysis technique we
assume $CP$ conservation and determine the decay asymmetry in $\CASMIN$,
combining $\Xi^+$ and $\Xi^-$ events.  We find $\alpha_{\Xi,\bar{\Xi}}
\alpha_{\LA,\bar{\LA}}=-0.291 \pm 0.021$ where the error is statistical.
The result is in good agreement with the world average $\alpha_{\Xi} \alpha_{\LA}
= -0.293 \pm 0.007 $~\cite{PDG}. 
\begin{figure}
\centerline{
\psfig{figure=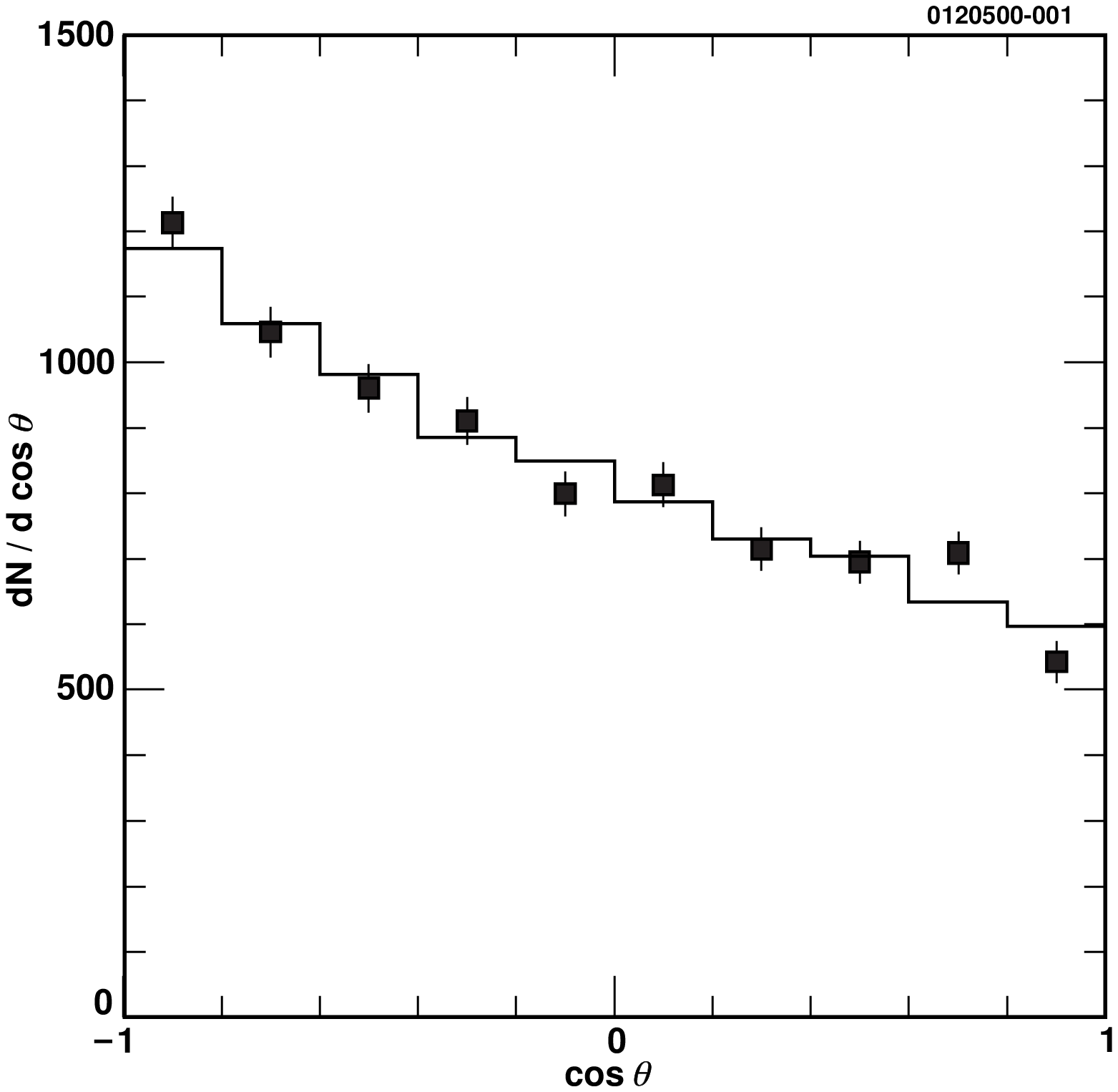,height=0.4\textwidth}
}
\caption{ 
The sideband subtracted angular distribution of the proton in the $\LA$ rest frame with
  respect to the direction of the $\LA$ in the $\Xi^-$ rest frame in
  $\Xi^- \to \LA \pi^+ , \LA \To p \pi^- $ for data 
  (points with
  error bars) and projection of the fit (line). Both charge conjugate states are included.}
\label{Xi14}
\end{figure}

We have considered the following sources of systematic uncertainties and
give our estimate of their magnitude in parentheses. The error
associated with finite MC statistics is estimated by varying the size of
the MC sample used in the fit $(1.6\%)$. The uncertainty from varying the
$\LA$ selection criteria and particle identification is
$(-6.3\%,+2.1\%)$.  The uncertainty associated with MC modeling of slow
pions from $\Xi$ and $\LA$ decay is obtained by varying the slow pion
reconstruction efficiency according to our understanding of the CLEO II
detector $(-0.0\%,+0.7\%)$.  The error due to finite interval size is
determined by varying the size of the interval for both signal and
background 
$(-0.5\%,+4.1\%)$. The uncertainty arising from incomplete knowledge of
the background shape in the sidebands is evaluated by varying the size
of the mass sideband region $(-1.2\%,+0.9\%)$. A binned fit is 
also performed and the result is consistent with the maximum likelihood fit.
This measurement is
insensitive to production polarization, ${\bf P_{\Xi}}$, and no systematic 
error has been included from this
source (see below).  Combining all sources of systematic errors in quadrature
the total systematic error is $(-6.6\%,+5.0\%)$ and $\alpha_{\Xi,\bar{\Xi}}
\alpha_{\LA,\bar{\LA}}= \SLOPEXA$.

\begin{figure}
\centerline{\psfig{figure=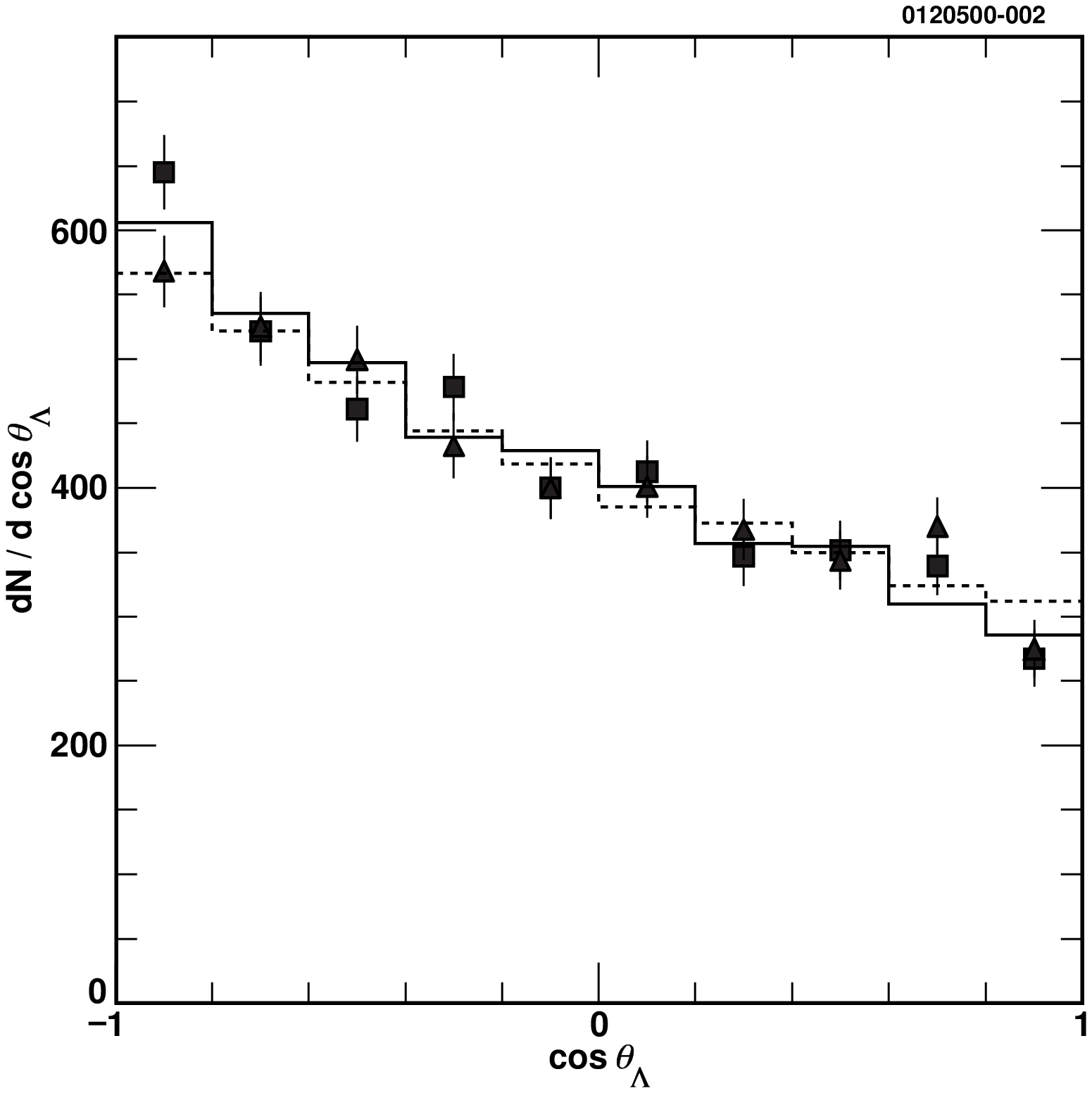,height=0.4\textwidth}}
\caption{
  The sideband subtracted angular distribution of the $p$ in the $\LA$ rest frame with
  respect to the direction of the $\LA$ in the $\Xi^-$ rest frame in
  $\Xi^- \to \LA \pi^+ , \LA \To p \pi^- $ for data (triangles with
  error bars) and projection of the fit (dashed line). The sideband subtracted angular
  distribution of the $\bar{p}$ in $\bar{\Xi}^+ \to \bar{\LA} \pi^+,
  \bar{\LA} \To \bar{p} \pi^-$ in data (squares with error bars) and
  projection of the fit (solid line).  }
\label{cpfit}
\end{figure}

To search for $CP$ violation the $\Xi$ candidates are sorted
into charge conjugate states. There are  $4204 \pm 75$ $\Xi^-$ decays and
$4200 \pm 75$ $\Xi^+$ decays.  Using the same fitting procedure as above,
and one MC sample for each charge conjugate state, we measure
$\alpha_{\Xi} \alpha_{\LA} = -0.275 \pm
0.030^{+0.019}_{-0.015}$ and $\alpha_{\bar{\Xi}} \alpha_{\bar{\LA}} =
-0.308 \pm 0.030^{+0.021}_{-0.016}$ respectively. Data and fit
projections are shown in Fig.~\ref{cpfit}.  From these results we
calculate ${\mathcal{A}}  = 0.057 \pm 0.064$
where the statistical error is estimated using the method in
~\cite{E871}.
Most systematic errors associated with $\Xi^-$ reconstruction apply
equally to $\Xi^+$ and cancel when determining the asymmetry
${\mathcal{A}}$.  If there is any $\Xi$ production polarization, it is
required to be normal to the production plane. The angular distribution
with respect to this plane will have opposite slopes for $\Xi^-$ and
$\bar{\Xi}^+$. However both the production plane and detector acceptance 
are uniformly distributed in the azimuthal angle $\phi$. Therefore
this result is insensitive to production
polarization.  If there is any difference between the
relative angular efficiencies of $\Xi^-$ and $\Xi^+$ that is not
modeled accurately by MC, this would lead to a biased value of ${\mathcal{A}}$.  
A momentum dependent asymmetry in
detection efficiencies for negative and positive soft pions could, in
principle, exist due to differing hadronic interaction cross sections of
$\pi^-$ and $\pi^+$ in the 3.5 cm radius CLEO II beryllium beam pipe.
We assume this effect to be absent for two reasons. (1) In this
analysis, typical decay lengths for $\Xi$ and $\LA$ 
are 6 cm and 11 cm respectively, therefore most tracks 
originate outside the beam pipe. (2) 
There is no evidence for a momentum dependent asymmetry in the CLEO II 
detection efficiency for tracks that originate inside the beampipe~\cite{D0CP}.

The $\LA \pi^- (\bar{\LA} \pi^+)$ mass sidebands consist mostly of
real $\LA$'s paired with random pions. The efficiency corrected angular
distribution of the sidebands in each charge conjugate state will be
the same if the MC accurately models the relative angular
efficiency of slow $\pi^+$ and $\pi^-$.  To study this precisely,
a high statistics sideband sample is obtained by relaxing
the
selection criteria and increasing
the width of the sideband regions. 
The fit, performed on $12365 \pm 111 (12353 \pm 111)$
events in the $\LA \pi^- (\bar{\LA}
\pi^+)$ sideband regions, yields $\alpha_{\Xi}
\alpha_{\LA} = 0.062 \pm 0.016 $ in the $\Xi^-$ sideband and $
\alpha_{\bar{\Xi}} \alpha_{\bar{\LA}} = 0.077 \pm 0.016 $ in the
$\bar{\Xi}^+$ sideband. The difference between these two values
$ -0.015 \pm 0.023$ indicates that there is no
statistically significant discrepancy in the modeling of the $\Xi^-$ and
$\bar{\Xi}^+$ detector angular acceptance in MC compared to data.
The error on the difference 
is taken as the systematic error on the $CP$ asymmetry parameter
measurement.  The result is
\begin{equation}
{\mathcal{A}}  = -0.057 \pm 0.064 \pm  0.039 \\
\end{equation}
There are three measurements of $A_{\LA}$~\cite{CHAUVAT}~\cite{TIXIER}~\cite{BARNES96}. 
Using the most precise:
$A_{\LA} = +0.013 \pm 0.022$~\cite{BARNES96} we obtain
$A_{\Xi}=-0.070 \pm 0.064 \pm 0.045$. Where the second error is systematic
and includes the error on $A_\LA$. 

It is possible to obtain a more precise value for $ {\mathcal{A}}$ by
combining our value for $\alpha_{\bar{\Xi}} \alpha_{\bar{\LA}}$ with the
PDG value of $\alpha_{\Xi} \alpha_{\LA}$ to obtain
${\mathcal{A}}  = -0.025^{+0.061}_{-0.056}$
and $A_{\Xi} = -0.038^{+0.065}_{-0.060}$ where the statistical and systematic
errors have been combined.

Finally, as this measurement of $\alpha_{\bar{\Xi}} \alpha_{\bar{\LA}}$
is significantly more precise than the only published value~\cite{STONE}, by combining with:
$\alpha_{\bar{\LA}}= -0.63 \pm 0.13$~\cite{TIXIER}, we obtain
$\alpha_{\bar{\Xi}}= 0.49 ^{+0.06}_{-0.05}\pm 0.10 $, where
the first error is the combined statistical and systematic error from
this study and the second error is from
$\alpha_{\bar{\LA}}$.

In conclusion, from a sample of approximately 4,000
$\Xi$ decays and similar number of $\bar{\Xi}$, we have measured $\alpha_{\bar{\Xi}}
\alpha_{\bar{\LA}} = -0.308 \pm 0.030 ^{+0.021}_{-0.016}$, and obtained
$\alpha_{\bar{\Xi}}= 0.49 ^{+0.06}_{-0.05}\pm 0.10 $.  
We
have searched for direct CP violation by measuring the asymmetry
parameter ${\mathcal{A}}$. 
We find no
evidence for $CP$ violation: ${\mathcal{A}} = \ACP$ and $A_{\Xi} =
-0.070 \pm 0.064 \pm 0.045$\footnote{As this 
paper was about to be
submitted, we became aware of K.B. Luk {\it et al.} hep-ex/0007030, 
July 2000. This paper measures $\alpha_{\bar{\Xi}} \alpha_{\bar{\LA}}$, 
${\mathcal{A}}$ and $A_{\Xi}$ with
greater precision than the values reported here. The two analyses are
consistent but employ different experimental techniques.}.


We gratefully acknowledge the effort of the CESR staff in providing us
with excellent luminosity and running conditions.  This work was
supported by the National Science Foundation, the U.S. Department of
Energy, Research Corporation, the Natural Sciences and Engineering
Research Council of Canada, the A.P. Sloan Foundation, the Swiss
National Science Foundation, and the Alexander von Humboldt Stiftung.


\begin{thebibliography}{99}
\bibitem{BRAN} In the MSM there is also a small contribution to $\epsilon$ from 
interference between mixing and decay amplitudes.
\end{thebibliography}
\end{document}